\documentclass[conference]{IEEEtran}

\usepackage{graphicx} 
\usepackage{booktabs} 
\usepackage[tbtags]{amsmath}
\usepackage{amsfonts}
\usepackage{amssymb}
\usepackage{dblfloatfix}
\usepackage[caption=false,font=footnotesize]{subfig}
\usepackage{color}
\usepackage{enumitem}
\usepackage{amsthm}
\usepackage{algorithmic}
\usepackage{multirow}
\usepackage{array}
\usepackage{cite}
\usepackage[normalem]{ulem}

\begin{document}
%
\title{Analysis of the Communication Traffic for Blockchain Synchronization of IoT Devices}

\author{\IEEEauthorblockN{Pietro Danzi, Anders Ellersgaard Kal{\o}r, \v{C}edomir Stefanovi\'c, Petar Popovski,\\}
\IEEEauthorblockA{Department of Electronic Systems, Aalborg University, Denmark \\
Email: \{pid,aek,cs,petarp\}@es.aau.dk }
}

\maketitle

\begin{abstract}

Blockchain is a technology uniquely suited to support massive number of transactions and smart contracts within the Internet of Things (IoT) ecosystem, thanks to the decentralized accounting mechanism.
In a blockchain network, the states of the accounts are stored and updated by the validator nodes, interconnected in a peer-to-peer fashion.
IoT devices are characterized by relatively low computing capabilities and low power consumption, as well as sporadic and low-bandwidth wireless connectivity. An IoT device connects to one or more validator nodes to observe or modify the state of the accounts.
In order to interact with the most recent state of accounts, a device needs to be synchronized with the blockchain copy stored by the validator nodes.
In this work, we describe general architectures and synchronization protocols that enable synchronization of the IoT endpoints to the blockchain, with different communication costs and security levels.
We model and analytically characterize the traffic generated by the synchronization protocols, and also investigate the power consumption and synchronization trade-off via numerical simulations.
To the best of our knowledge, this is the first study that rigorously models the role of wireless connectivity in blockchain-powered IoT systems. 
\end{abstract}


\section{Introduction}
The blockchain protocols have recently attracted enormous interest from researchers and industry, appearing to be a promising, but still not mature, technology.
This protocol family stems from the Bitcoin specification \cite{nakamoto2008bitcoin}, proposing a cryptographic-based solution to the problem of keeping consistent copies of a distributed database in a permission-less network.
The idea has been extended by Ethereum protocol \cite{wood2014ethereum}, where the database that is used to keep track of the states of accounts was augmented by a scripting functionality that permits any computer in the network to reconstruct the accounts' state by simply reading the database.
However, the database is hard to tamper with, as its modification requires the investment of computational resources.

The blockchain protocols offer interesting applications for the IoT ecosystem \cite{dorri2017towards}, mainly seen in the possibility of stipulating \emph{smart contracts} \cite{christidis2016blockchains}. 
A smart contract is an account, whose state can be read and modified by any device according to predefined functions.
If the state contains credits, a smart contract can regulate economic transactions, e.g., \emph{micro-payments}, among devices.
For instance, a smart contract can be used to mimic the presence of a central authority in a decentralized smart energy application, where the contract is stored and updated by devices themselves~\cite{danzi2017distributed}.
However, in the IoT context, the modest storage and processing capabilities of the devices may require the presence of an external infrastructure to store the smart contracts.

\begin{figure}[!t]
\centering
  \includegraphics[width=0.75\columnwidth]{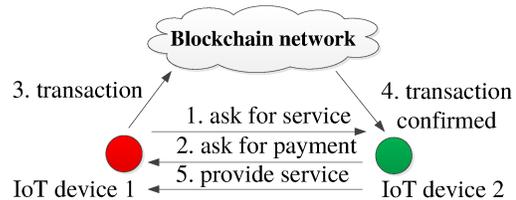}
  
    \caption{Information exchange for the micro-payments.}
    \label{fig:payment}
\end{figure}

Fig.~\ref{fig:payment} illustrates the steps of a credit transaction regulated by smart contract. 
Two IoT devices are connected to the peer-to-peer blockchain network, whose nodes store the smart contract, maintain the blockchain and update it with new information blocks.
A device can request the blockchain network to update the smart contract state, e.g., to attribute some credit to another device (step 3 in Fig.~\ref{fig:payment}).
The other devices are informed of this update when they receive a new block containing the updated state (step 4 in Fig.~\ref{fig:payment}).
Finally, the recipient device can verify the new state and release a service to the sender (step 5 in Fig.~\ref{fig:payment}).
The mechanism enables the exchange of credit in absence of mutual trust, since the economic transaction is certified by a third party, i.e., the blockchain network, and hardly reversible \cite{narayanan2016bitcoin}.
The objective is to assure that each device locally observes the same state of the smart contract, such that the key element in the system design is how each device stays synchronized with the most recent version of the blockchain.

In this paper, we characterize the traffic between the blockchain network and the IoT devices.
The critical problems in this context are: (i) the low computational and (ii) storage resources that prevent IoT devices to act as blockchain validators, and (iii) limited communication resources.
In fact, smart meters, and IoT devices in general, access the network via low-bandwidth wireless links. 
In addition, their connectivity is sporadic, e.g., due to energy saving policies, or limited access to channel resources.
In this work, we establish two possible protocols for interacting with the blockchain network. Each protocol has different communication requirements and provides a different security level.
We then propose a traffic model that can be used to extract measures of interest, such as the time needed for the blockchain synchronization, 
or the minimal bandwidth requirements to stay synchronized.
We analyze the probability of keeping the IoT devices synchronized with the blockchain and investigate trade-offs among duty cycle (i.e., sleeping duration), goodput offered by the wireless link, and blockchain parameters (i.e., the block generation frequency and its size).
The key contribution is the explicit modeling of the impact that wireless connectivity has on the blockchain synchronization and smart contract execution. 

The rest of the paper is organized as follows.
Section~\ref{sec:protocol} introduces the main concepts of a blockchain protocol.
Section~\ref{sec:model} presents the system model.
Section~\ref{sec:arch} elaborates the synchronization protocols under consideration, whose analysis is perfomed in Section~\ref{ref:analysis}.
Section~\ref{sec:results} presents the numerical results.
Section~\ref{sec:conclusion} concludes this paper.

\section{Blockchain protocol}\label{sec:protocol}

In this section, we focus on the blockchain concepts that are relevant for this work; the reader is referred to \cite{narayanan2016bitcoin} for a detailed overview.
A blockchain is a concatenated list of blocks, stored in multiple copies by the nodes of the blockchain network. 
Part of the nodes act as \emph{validators}, i.e., they append new blocks to their local copy and also send them to the rest of the nodes.
The distribution of new blocks happens in a peer-to-peer fashion.
When a block is received, it is validated before it can be appended to the local list.
In order to be perform validation, a node has to invest resources that are referred to as Proof of Works (PoWs); PoWs are typically computational resources needed to solve cryptographic-based problems. 
In case that different subsets of nodes have different local blockchains with the same number of blocks (e.g., caused by a communication network partition) the contention is solved in a consensus-based manner.
At any time, the validators trust the longest known blockchain in the network.

\subsubsection{Data structures}\label{sec:ds}

A blockchain block is composed of a header and a body.
The body contains the list of transactions that are verified by the block, and has a maximum size.
Each transaction is represented by a data structure, composed by a set of fields, defined in the protocol specifications \cite{nakamoto2008bitcoin},\cite{wood2014ethereum}.
The header typically contains a pointer to the previous block, a solution to the PoW, the block creation time, and one or more roots of Merkle trees~\cite{merkle1987digital} that are used as Proof of Inclusion (PoI) of the transactions in the block, see Sec.~\ref{sec:poi}.

\subsubsection{Proof of Work (PoW)}

The PoW algorithm prevents uncontrolled generation of blocks by the validators, and guarantees that their local blockchain copies are consistent.
The algorithm is run locally by all validators, with the aim to find a solution to a cryptographic puzzle~\cite{nakamoto2008bitcoin}.
The difficulty of the puzzle is tuned to keep the expected block generation time constant, typically of the order of tens of seconds~\cite{narayanan2016bitcoin}.

\subsubsection{Proof of Inclusion (PoI)}\label{sec:poi}

The Merkle tree roots contained in the header can be used to prove the inclusion of transactions in the block without the need to download the entire block body.
Specifically, cryptographic hash-values of the transactions that are included in a block are stored in a binary Merkle tree~\cite{nakamoto2008bitcoin}.
Therefore, any node in possession of the block header can verify a transaction from another node by requesting the transaction's data structure, together with the Merkle-tree nodes that are required to reconstruct the tree root.
If the root matches the one included in the block header, then the transaction is included in the block.
In Ethereum, a similar mechanism is used to prove the state of accounts, e.g., smart contracts~\cite{wood2014ethereum}.

\section{System model}\label{sec:model}

\subsubsection{Blockchain network} The blockchain network relies on peer-to-peer connections between nodes, including validators that produce information blocks and store the entire blockchain list.
We assume that validators are interconnected by an ideal communication network.
Furthermore, validators are assumed to be honest, and, at any time, they store the same copy of the blockchain.
The chain length at time $t$, i.e., number of generated blocks, is characterized by the homogeneous Poisson process $h(t)$ with inter-arrival frequency $\lambda_\mathrm{B}$.

\subsubsection{The IoT device} We consider a generic IoT device $u$, with limited storage and energy.
For these reasons, it is not feasible for $u$ itself to validate the blockchain, but instead it relies on information received from the blockchain network.
The analysis performed in the paper focuses on the case where $u$ is interested in verifying that a transaction has been included in the blockchain (step 4 in Fig.~\ref{fig:payment}), rather than on the transaction creation.
This is motivated by the fact that the creation of transactions is decoupled from their verification, where communication demands of sending a transaction to the blockchain (step 3 in Fig.~\ref{fig:payment}) are negligible compared to the reception of blocks.

In order to verify transactions, $u$ stores a local copy of the (global) blockchain.
We denote the length of the local blockchain by $h_u(t)\leq h(t)$, which is never longer than the global, since $u$ is not generating blocks.
The process $H(t) = h(t) - h_u(t)$ describes the difference in the number of blocks between the global and local copy (i.e., how many blocks $u$ is delayed with respect to the global blockchain) at time $t$.
The delay in the local copy is caused by blocks created while $u$ is sleeping to save energy.

Device $u$ is connected to the blockchain network by a wireless transceiver that connects $u$ with $N \ge 1$ blockchain-network (BN) nodes.
The connectivity of $u$ is modelled by the two-state process $S(t)$ with states $\mathrm{S}_0$ and $\mathrm{S}_1$.
In state $\mathrm{S}_0$, $u$ is connected and the synchronization protocol is triggered by the creation of new blocks.
In state $\mathrm{S}_1$, $u$ is sleeping and does not receive block updates or execute the protocol.
Furthermore, a transition from state $\mathrm{S}_1$ to $\mathrm{S}_0$ triggers a protocol execution whose duration is random; the details are given in Sec.~\ref{sec:arch}.
Additionally, an inadequate communication link capacity may cause the execution of a synchronization protocol to last longer than the block creation period, requiring multiple executions in order to achieve synchronization.

\begin{figure}[!t]
\centering

{\includegraphics[width=\columnwidth]{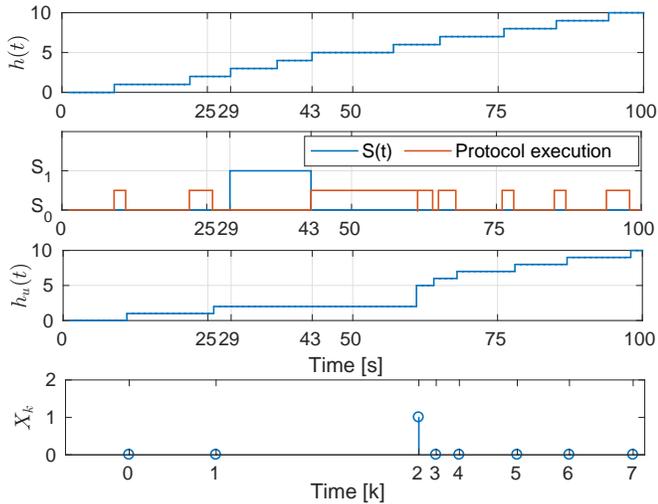}}

    \caption{Relationship of block creation process, IoT device state, and delay of local blockchain.}
		\label{fig:time_organization}
\end{figure}

We are interested in characterizing $H(t)$ after the execution of the synchronization protocol. To this end, we define the discrete-time process, $X_k$, which samples $H(t)$ at the end of the protocol executions.
A realization of the introduced processes is illustrated in Fig. \ref{fig:time_organization} with $\lambda_\mathrm{B}=0.1 \, \mathrm{s}^{-1}$.
Observe that $u$ is sleeping in interval $t=[29,43) \, \mathrm{s}$, causing a delay of the local blockchain copy at $k = 2$.
The blockchain consistency is achieved after the next protocol execution.

\subsubsection{Connectivity model} The connectivity process $S(t)$ of a device can be extremely variable and depends on the type of device.
For instance, a smart meter may wake up periodically for short time intervals every 15--30 minutes~\cite{eibl2015influence}.
At the same time, a smartphone may be turned off during the night hours, while being connected the rest of the day.
In this work, we consider a model where $u$ goes to sleep after protocol execution with probability $p_{\mathrm{s}}$, and then sleeps for a deterministic duration $t_\mathrm{s}$.

\subsubsection{Communication link} We model $u$'s connectivity to the blockchain network as a stop-and-wait retransmission protocol with instant and error-free feedback and independent packet error probabilities $p_{\mathrm{e,DL}}$ in the downlink and $p_{\mathrm{e,UL}}$ in the uplink.
For simplicity, we assume that the error probabilities are independent of packet length and that the sender has unlimited retransmission attempts (i.e., its packets are eventually delivered).
Under this scheme, the number of transmission attempts follows a geometric distribution.
As a result, the expected goodputs in the downlink and uplink are $g_{\mathrm{DL}}=g_{\mathrm{DL}}^\mathrm{M} (1-p_{\mathrm{e,DL}})$ and  $g_{\mathrm{UL}}=g_{\mathrm{UL}}^\mathrm{M} (1-p_{\mathrm{e,UL}})$, where $g_{\mathrm{DL}}^M$ and $g_{\mathrm{UL}}^M$ are the downlink and uplink bit-rates, respectively.
Finally, we also assume that there is an initial connection establishment and authentication phase of deterministic duration, $t_{\mathrm{c}}$.

\begin{figure}[!tb]
\centering
  \subfloat[Point-to-multipoint.]{\includegraphics[width=0.65\columnwidth]{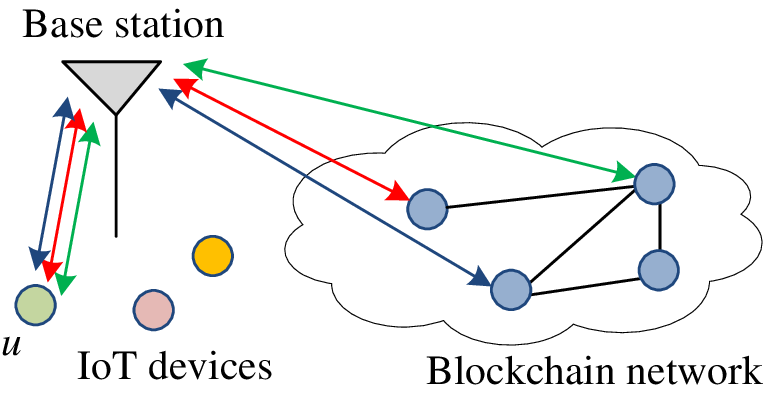}}
  
   \subfloat[Point-to-point.]{ \includegraphics[width=0.65\columnwidth]{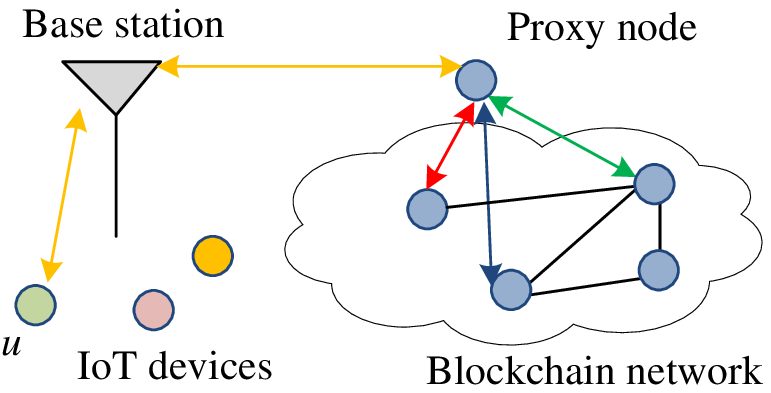}}
    \caption{The considered system architectures.}
    \label{fig:architecture}
\end{figure}

The connection to the blockchain network follows a point-to-multipoint topology when $N > 1$; when $N=1$, $u$ connects to a single node that acts as a proxy, see Fig.~\ref{fig:architecture}.
In the latter case, the proxy needs to be fully trusted, since it can modify the information forwarded to the IoT device~\cite{heilman2015eclipse}.
This configuration is not investigated in this paper, as it prevents the implementation of a fully decentralized architecture.

\section{Blockchain synchronization protocols}\label{sec:arch}

The task of a blockchain synchronization protocol is to keep $u$ constantly updated with the current state of the blockchain.
Motivated by the fact that the protocol steps are not explicitly described by the most popular blockchain specifications (e.g.\ \cite{nakamoto2008bitcoin,wood2014ethereum}), but left to the implementation, we define two conceptual synchronization protocols: P1 and P2.
They model the minimal amount of information that needs to be exchanged, while neglecting additional features included in specific blockchain implementations.

P1 is the simplest protocol variant, in which $u$ receives complete blocks from BN nodes to which is connected, and locally verifies the validity of the PoW solution and the contained transactions.
This configuration, adopted in~\cite{danzi2017distributed}, provides the maximum possible level of security: $u$ needs not trust the individual BN nodes and is designated as a \emph{full node}.
However, this is hardly a feasible solution for IoT due to the high storage and computational requirements, since $u$ needs to store the complete blockchain and to check all transactions.

Protocol P2 provides a simplified architecture where $u$, acting as a \emph{light node}, only receives the block headers from BN nodes by default.
Furthermore, $u$ defines a list of events that it is interested in observing, i.e. modifications to the blockchain, and this list is known by BN nodes.\footnote{In efficient implementations, the list is constructed via a Bloom filter that is matched when the events happen~\cite{nakamoto2008bitcoin}.}
Examples of events of interest are the modification of the state of a smart contract, or transactions from/to a particular address.
Upon the observation of the events, BN nodes send to $u$ the part of information that has changed, together with the corresponding Merkle tree that serves as a PoI, see Sec.~\ref{sec:poi}.
We assume that an event of interest is independently observed in each block period with probability $p_{\mathrm{M}}$, which can be tuned by $u$ by changing the set of events of interests.

To simplify the scenario, we assume that $u$ can only interact with one BN node at a time (no concurrent transfers), and that there is a single event of interest.
In case that the event is observed while $u$ is sleeping, the current state of the blockchain becomes reconstructed based on the header of the most recent block.
If $u$ is synchronized and a new block becomes generated, the protocol execution starts immediately.
If $u$ is not synchronized after the protocol execution ends, due to the generation of a new block during the execution, a new protocol execution is started immediately afterwards.
When an execution starts, it does not stop until it is completed, and $u$ stays awake during the execution.
In the following, we describe possible implementations for P1 and P2.

\subsection{Protocol P1}

\begin{figure}[!t]
\centering

{\includegraphics[width=0.75\columnwidth]{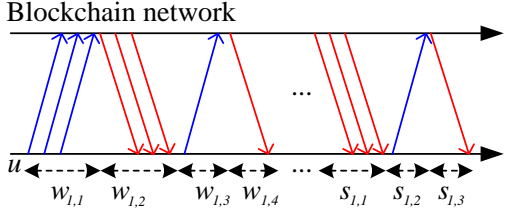}}

    \caption{Message exchanges in P1. Downlink is denoted in red, uplink in blue.}
		\label{fig:p1mess}
\end{figure}

The information exchange required by P1 depends on whether $u$ was sleeping or being idle prior to the protocol execution.
Specifically, when $u$ wakes up, i.e., makes a transition $\mathrm{S}_1 \rightarrow \mathrm{S}_0$, the following sequence is executed:
\begin{itemize}[leftmargin=10mm]
\item[($\mathrm{w}_{1,1}$)] $u$ request new blocks from each of the BN nodes via a message of length $l_\mathrm{n}$ bits, which contains information about the highest locally known block $h_\mathrm{u} (t)$.
\item[($\mathrm{w}_{1,2}$)] $u$ receives a response of length $l_\mathrm{r}$ bits, from each BN node, which contains the information of their current blockchain length $h(t)$.
\item[($\mathrm{w}_{1,3}$)] If $ h(t) - h_u(t) = n > 0$ (i.e., $X_\mathrm{k} > 0 $), $u$ selects one random BN node and sends a message of length $l_{\mathrm{n}}$ bits, requesting the missing $n$ blocks.
\item[($\mathrm{w}_{1,4}$)] The selected BN node sends the blocks. 
Since there are $n$ blocks, and each block is of length $l_\mathrm{b}$ bits, the total length of the message is $n \cdot l_\mathrm{b}$ bits.
\item[($\mathrm{w}_{1,5}$)] Steps ($\mathrm{w}_{1,3}$)--($\mathrm{w}_{1,4}$) are repeated until synchronization.
\end{itemize}

In the second case, $u$ is connected and receives notifications of new blocks. The steps are:
\begin{itemize}[leftmargin=10mm]
\item[($\mathrm{s}_{1,1}$)] $u$ receives a notification of the new block from each of the BN nodes, where the length of the notification message is $l_\mathrm{r}$ bits. Starting from the first notification, it waits $t_\mathrm{w}$ seconds, to receive all the notifications.
\item[($\mathrm{s}_{1,2}$)] $u$ selects one BN node and requests the missing block.
\item[($\mathrm{s}_{1,3}$)] The selected BN node sends the block.
\end{itemize}

The message exchanges are depicted in Fig.~\ref{fig:p1mess}.

\subsection{Protocol P2}

The operation of P2 are similar to P1, but the message sizes in downlink are remarkably smaller. When $u$ wakes up:
\begin{itemize}[leftmargin=10mm]
\item[($\mathrm{w}_{2,1}$)] $u$ requests of new block to each of the BN nodes, via a message of length $l_\mathrm{n}$ bits.
\item[($\mathrm{w}_{2,2}$)] $u$ receives a response of length $l_\mathrm{r}$ bits from each of the BN nodes, that contains the list of new blocks available.
\item[($\mathrm{w}_{2,3}$)] $u$ selects one BN node and requests the missing $n$ block headers.
\item[($\mathrm{w}_{2,4}$)] The selected BN node sends the headers. Since each header is $l_\mathrm{h}$ bits long, the message length is $n \cdot l_\mathrm{h}$ bits.
\item[($\mathrm{w}_{2,5}$)] Steps ($\mathrm{w}_{2,3}$)--($\mathrm{w}_{2,4}$) are repeated until synchronization.
\end{itemize}

If $u$ is connected and synchronized:
\begin{itemize}[leftmargin=10mm]
\item[($\mathrm{s}_{2,1}$)] $u$ receives a notification of new block of length $l_\mathrm{r}$ bits, from each of the BN nodes. Upon the reception of the first notification, $u$ waits for $t_\mathrm{w}$ seconds.
\item[($\mathrm{s}_{2,2}$)] $u$ selects one BN node and requests the missing header.
\item[($\mathrm{s}_{2,3}$)] The selected BN node sends the header.
\item[($\mathrm{s}_{2,4}$)] If an event of interest is observed, the BN node also sends the data structure containing the information of interest, which is of length $l_\mathrm{i}$ bith, together with the PoI that is of length $l_\mathrm{PoI}$ bits.
\end{itemize}

\section{Analysis}\label{ref:analysis}
\begin{figure}[!t]
\centering

{\includegraphics[width=0.86\columnwidth]{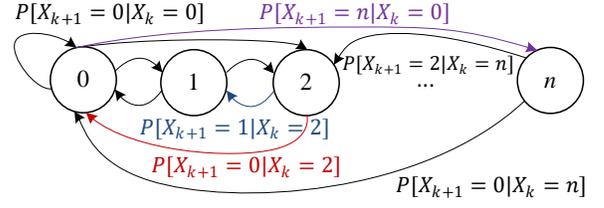}}

    \caption{The modelled process and examples of state transitions.}
		\label{fig:transitions}
\end{figure}

Observe that the adopted sampling makes $X_k$ an irreducible and positive recurrent Markov process. 
The goal of the analysis is to find its transition probabilities (see Fig. \ref{fig:transitions}), which depend on the adopted protocol and which are derived in the sequel.

The probability of transitioning from state $X_k=n$ to $X_{k+1}=m$, where the state corresponds to the number of missing blocks in the local copy of the blockchain, depends on whether $u$ is sleeping between $X_k$ and $X_{k+1}$:
\begin{equation}\label{eq:transprob}
P[ X_{k+1} = m | X_{k} = n ] = (1-p_\mathrm{s}) \cdot p_{\mathrm{S}_0}(m|n) + p_\mathrm{s} \cdot p_{\mathrm{S}_1}(m|n).
\end{equation}
The first term, corresponding to the case that $u$ stays awake between two protocol executions, is the probability that $m$ blocks arrive during the protocol execution time $t_\mathrm{p}$:
\begin{align}\label{eq:transprob1}
p_{\mathrm{S}_0}(m|n) &= \int_{0}^\infty p_{\mathrm{S}_0}(t_\mathrm{p}|n)\frac{(\lambda_\mathrm{B} t_\mathrm{p})^{m}}{m!} e^{-\lambda_\mathrm{B} t_\mathrm{p}}dt_\mathrm{p}
\end{align}
where we use the fact that blocks are generated with Poisson arrivals with parameter $\lambda_{\mathrm{B}}$.
For the case where $u$ sleeps between $X_k$ and $X_{k+1}$, the protocol execution time depends on $n$ and the number of blocks accumulated during sleep, denoted by $q$. By marginalizing over $q$ we obtain
\begin{align}\label{eq:transprob2}
  \begin{split}
&p_{\mathrm{S}_1}(m | n) = \sum_{q=0}^{\infty} p(m|n,q)p(q)\\
&= \sum_{q=0}^{\infty} \left(\int_0^\infty p_{\mathrm{S}_1}(t_\mathrm{p}|n,q)\frac{(\lambda_\mathrm{B} t_\mathrm{p})^{m}}{m!} e^{-\lambda_\mathrm{B} t_\mathrm{p}}dt_\mathrm{p}\right)
\frac{(\lambda_\mathrm{B} t_\mathrm{s})^{q}}{q!} e^{-\lambda_\mathrm{B} t_\mathrm{s}}.
\end{split}
\end{align}
Note that blockchain protocols might specify a limit on the number of blocks that can be transferred with one protocol execution, which can be easily incorporated in the proposed model.
Also, here we assume the reconnection time (i.e., the time for discovery of BN nodes and connection negotiation) to the blockchain network to be negligible and note that the model can be easily extended by including reconnection time into $t_c$.
We continue by deriving $p_{\mathrm{S}_0}(t_\mathrm{p}|n)$ and $p_{\mathrm{S}_1}(t_\mathrm{p}|n)$ for the proposed protocols.

\begin{figure}[!t]
\centering

{\includegraphics[width=0.7\columnwidth]{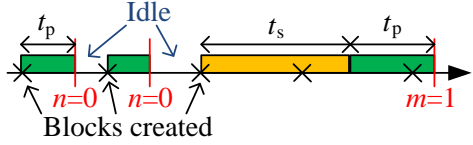}}

    \caption{Visualization of protocol execution and block creation process.}
		\label{fig:conditional}
\end{figure}

\subsection{Protocol P1}
\label{sec:P1}
We denote the expected time in order to successfully deliver a generic protocol message $x$ by $\Delta t_x$.
In case that $u$ is awake and synchronized, i.e. $n=0$, the protocol execution time is
\begin{align*}
  \begin{split}
  t_\mathrm{s,1} &= \Delta t_{\mathrm{s}_{1,1}} + \Delta t_{\mathrm{s}_{1,2}} + \Delta t_{\mathrm{s}_{1,3}}
  \\&=g_\mathrm{DL}^{-1} l_\mathrm{r} + t_\mathrm{w} + g_\mathrm{UL}^{-1}l_\mathrm{r} + g_\mathrm{DL}^{-1}l_\mathrm{b}.
\end{split}
\end{align*}
If $u$ is awake and not synchronized, the duration of the execution is a function of the number of missing blocks $n$:
\begin{align*}
  \begin{split}
    t_{\mathrm{w},1}(n) &= \Delta t_{\mathrm{w}_{1,1}} + \Delta t_{\mathrm{w}_{1,2}} + \Delta t_{\mathrm{w}_{1,3}} + \Delta t_{\mathrm{w}_{1,4}}\\
    &=g_\mathrm{UL}^{-1} N l_\mathrm{n} + g_\mathrm{DL}^{-1} N l_r
    + g_\mathrm{UL}^{-1} l_\mathrm{n} + g_\mathrm{DL}^{-1}  n l_\mathrm{b}.
  \end{split}
\end{align*}

The distributions of the protocol execution time $t_\mathrm{p}$ are then
\begin{align}
  p_{\mathrm{S}_0}(t_\mathrm{p}|n) &=
    \begin{cases}
    \delta\left(t_{\mathrm{s},1}\right) & n=0\\
    \delta\left(t_{\mathrm{w},1}(n)\right) & \text{otherwise,}
  \end{cases}\\
  p_{\mathrm{S}_1}(t_\mathrm{p}|n,q)&=\delta\left(t_\mathrm{c} + t_{\mathrm{w},1}(n+ q )\right),
\end{align}
where $\delta(\cdot)$ is the Dirac delta function. 
If $u$ was sleeping prior to the protocol execution, the connection time $t_\mathrm{c}$ should be included as well: we define $t_\mathrm{a}(k)=t_\mathrm{c} + t_{\mathrm{w},1}(k)$ and insert it into \eqref{eq:transprob1} and \eqref{eq:transprob2} to obtain
\begin{align}
  p_{\mathrm{S}_0}(m|n) &=
  \begin{cases}
    \frac{(\lambda_{\mathrm{B}} t_{\mathrm{s},1})^{m}}{m!} e^{-\lambda_{\mathrm{B}} t_{\mathrm{s},1}} & n=0\\
    \frac{(\lambda_{\mathrm{B}} t_{\mathrm{w},1}(n))^{m}}{m!} e^{-\lambda_{\mathrm{B}} t_{\mathrm{w},1}(n)} & \text{otherwise},
  \end{cases}\\
  \nonumber p_{\mathrm{S}_1}(m|n) &= \\ \sum_{q=0}^{\infty} & \left(\frac{(\lambda_{\mathrm{B}} t_{\mathrm{a}}(n+q))^{m}}{m!} e^{-\lambda_{\mathrm{B}} t_\mathrm{a}(n+q)}\right)
\frac{(\lambda_{\mathrm{B}} t_\mathrm{s})^{q}}{q!} e^{-\lambda_{\mathrm{B}} t_\mathrm{s}}.
\end{align}

\subsection{Protocol P2}

The analysis is similar to the one for P1, where we substitute $l_\mathrm{b}$ with $l_\mathrm{h}$ in all the equations, as in this case $u$ downloads only block headers. Hence, the time required to download the header in the synchronized state is:
\begin{align*}\nonumber
\begin{split}
t_\mathrm{s,2} &= \Delta t_{\mathrm{s}_{2,1}} + \Delta t_{\mathrm{s}_{2,2}} + \Delta t_{\mathrm{s}_{2,3}} + \Delta t_{\mathrm{s}_{2,4}} \\ &=
g_\mathrm{DL}^{-1} l_\mathrm{r} + t_\mathrm{w} + g_\mathrm{UL}^{-1} l_\mathrm{r} + g_\mathrm{DL}^{-1} l_\mathrm{h}.
\end{split}
\end{align*}

Another difference to P1 is seen in step $(\mathrm{s}_{\mathrm{2, 4}})$ that refers to the case when $u$ is awake and synchronized and an event of interest happens.
This triggers the transmission of the information modified on the blockchain along with the PoI, which takes $g_\mathrm{DL}^{-1} (l_\mathrm{i} + l_\mathrm{PoI})$ seconds.
The overall time to synchronize and receive the information modified on the blockchain, i.e., to execute steps $(\mathrm{s}_{\mathrm{2, 1}})$-$(\mathrm{s}_{\mathrm{2, 4}})$, is:
\begin{equation}\nonumber
t_\mathrm{sm} = t_\mathrm{s,2} + g_\mathrm{DL}^{-1} (l_\mathrm{i} + l_\mathrm{PoI}).
\end{equation}

When $u$ is not synchronized, the time required to download the headers in P2 is
\begin{equation}\nonumber
t_\mathrm{w,2}(n) =g_\mathrm{UL}^{-1} N l_\mathrm{n} + g_\mathrm{DL}^{-1} N l_r + g_\mathrm{UL}^{-1} l_\mathrm{n} + g_\mathrm{DL}^{-1}  n l_\mathrm{h}.
\end{equation}
We also define $t_\mathrm{a,2}(n) = t_\mathrm{w,2}(n) + t_\mathrm{c}$, similarly to $t_\mathrm{a,1}$ in Sec.~\ref{sec:P1}, which takes into account the reconnection time.
If at least one event of interest happened in one of the blocks, the protocol also transfers the corresponding information, and lasts $t_\mathrm{as} = t_\mathrm{a,2} + g_\mathrm{DL}^{-1} (l_\mathrm{i} + l_\mathrm{PoI})$ seconds.

The probability that a protocol execution ends with $m$ missing blocks with respect to the global blockchain is conditioned on the event that an event of interest is observed, see \eqref{eq:b1} and \eqref{eq:bloom}.
\begin{figure*}[!hbt]
\begin{align}
  \label{eq:b1} & p_{\mathrm{S}_0}(m|n) = \begin{cases}
    \frac{(\lambda_{\mathrm{B}} t_\mathrm{sm})^{m}}{m!} e^{-\lambda_{\mathrm{B}} t_\mathrm{sm}} p_\mathrm{M} + \frac{(\lambda_{\mathrm{B}} t_\mathrm{s,2})^{m}}{m!} e^{-\lambda_{\mathrm{B}} t_\mathrm{s,2}} (1-p_\mathrm{M}) & n=0\\
    \frac{(\lambda_{\mathrm{B}} t_{\mathrm{w},2}(n))^{m}}{m!} e^{-\lambda_{\mathrm{B}} t_{\mathrm{w},2}(n)} & \text{otherwise}
  \end{cases}\\
  \label{eq:bloom} & p_{\mathrm{S}_1}(m|n) = \sum_{q=0}^{\infty} \left( \frac{(\lambda_{\mathrm{B}} t_\mathrm{as})^{m}}{m!} e^{-\lambda_{\mathrm{B}} t_\mathrm{as}} \frac{(\lambda_{\mathrm{B}} t_\mathrm{s})^{q}}{q!} e^{-\lambda_{\mathrm{B}} t_\mathrm{s}} \cdot (1-(1-p_\mathrm{M})^q) + \frac{(\lambda_{\mathrm{B}} t_\mathrm{a,2})^{m}}{m!} e^{-\lambda_{\mathrm{B}} t_\mathrm{a,2}} \frac{(\lambda_{\mathrm{B}} t_\mathrm{s})^{q}}{q!} e^{-\lambda_{\mathrm{B}} t_\mathrm{s}} \cdot (1-p_\mathrm{M})^q \right)
\end{align}
\end{figure*}

\subsection{Average idle time}

The presented analysis can be used to find several parameters of interest.
As an example, here we show how to obtain the average time spent in idle state, when $u$ is not sleeping and not communicating (because of being synchronized), see Fig.~\ref{fig:conditional}.
We indicate the sojourn time in state $k$ as $Z_k$, then we have $Z_0 = e^{-\lambda_\mathrm{B} (t-t_\mathrm{p})}$ and $Z_k = \delta(t_\mathrm{p})$ for $k > 0$.
The set of random variables $\{ X_k, Z_k \}$ defines a Markov renewal process, used to compute the average time spent in idle state, $\pi_0 \lambda_{\mathrm{B}}^{-1} (1 - p_{\mathrm{s}})$, where $\pi_k$ is the stationary probability of $X_k$.

\section{Numerical results}\label{sec:results}
\begin{table}[]
\centering
\caption{Blockchain parameters.}
\label{tab:bcparam}
\begin{tabular}{|l|l|l|l|}
\hline
\textbf{Parameter} & \textbf{Value} & \textbf{Parameter} & \textbf{Value}  \\ \hline
$\lambda_\mathrm{B}$              & 1/12 $s^{-1}$          & $t_\mathrm{w}$ & 0.5 s  \\ \hline
$l_\mathrm{b}$              & 40 kbit       & $l_\mathrm{PoI}$ & 1536 bit \\ \hline
$l_\mathrm{n} = l_\mathrm{r} = l_\mathrm{h}$  & 800 bit    & $l_\mathrm{i}$ &  1000  bit \\ \hline
\end{tabular}
\end{table}

\begin{table}[]
\centering
\caption{Communication parameters.}
\label{tab:comm}
\begin{tabular}{|l|l|l|l|}
\hline
\textbf{Technology} & $g_\mathrm{UL}^\mathrm{M}$ & $g_\mathrm{DL}^\mathrm{M}$ & $t_\mathrm{c}$ \\ \hline
A          & 1 Mbps             & 1 Mbps             & 100 ms        \\ \hline
B             & 100 kbps             & 150 kbps            & 1 s           \\ \hline
\end{tabular}
\end{table}

In this section, we provide numerical results based on typical values of blockchain networks.
The blockchain parameters, i.e., block period and messages length, are extracted from Ethereum protocol statistics and listed in Table~\ref{tab:bcparam}.
We assume that $u$ is connected to $N=6$ BN nodes.
We evaluate the use of two wireless technologies, named A and B and characterized by different capabilities, see Table \ref{tab:comm}; the practical instances of technologies A and B are LTE Cat M1, and Bluetooth Low Energy, respectively, both oriented towards supporting IoT traffic.

\begin{figure}[!tb]
\centering
  \includegraphics[width=0.8\columnwidth]{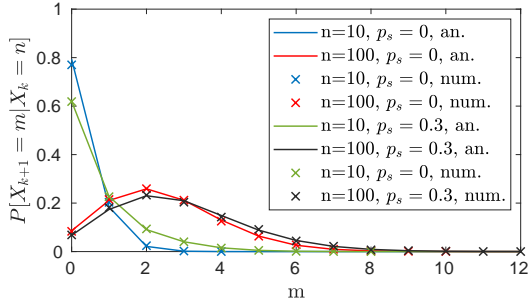}
    \caption{Comparison of analytical (an.) and numerical (num.) transition probabilities.}
    \label{fig:validation}
\end{figure}

First, we verify the validity of the analytical model, by comparing it with numerical simulation of the described processes, implemented in MATLAB.
Fig.~\ref{fig:validation} compares transition probabilities obtained from \eqref{eq:transprob} and from the numerical simulation for technology B and protocol P1, showing that the proposed model, although being simple, succeeds in capturing all the necessary details.
We note that investigations performed for protocol P2 and/or technology A produce similar results, which are omitted due to the space constraints.

\begin{figure}[!tb]
\centering
  \includegraphics[width=0.9\columnwidth]{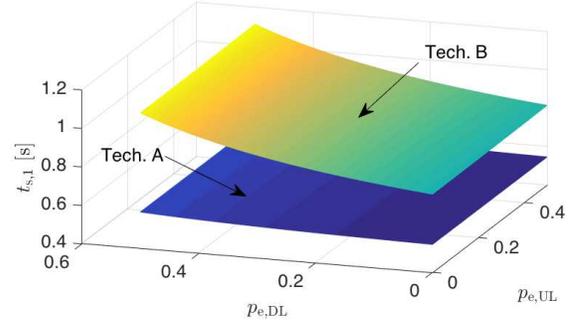}
    \caption{Duration of $t_\mathrm{s,1}$ for the considered technologies.}
    \label{fig:duration_ts1}
\end{figure}

In Fig.~\ref{fig:duration_ts1}, we show the duration of $t_\mathrm{s,1}$, i.e., time to execute P1 when $u$ is synchronized, for different values of $p_\mathrm{e,DL}$, $p_\mathrm{e,UL}$: we can observe the difference due to the different characteristics of the considered wireless interfaces.
The investigations performed for $t_\mathrm{a,1}$, $t_\mathrm{wc,1}$, $t_\mathrm{s,2}$, $t_\mathrm{sm}$, $t_\mathrm{as}$ show analogous results (here omitted) to the ones presented in Fig.~\ref{fig:duration_ts1}.

\begin{figure*}[!tb]
\centering
  \subfloat[Impact of sleeping time $t_\mathrm{s}$ and packet-error probability in the downlink $p_\mathrm{e,DL}$.]{\includegraphics[width=0.63\columnwidth]{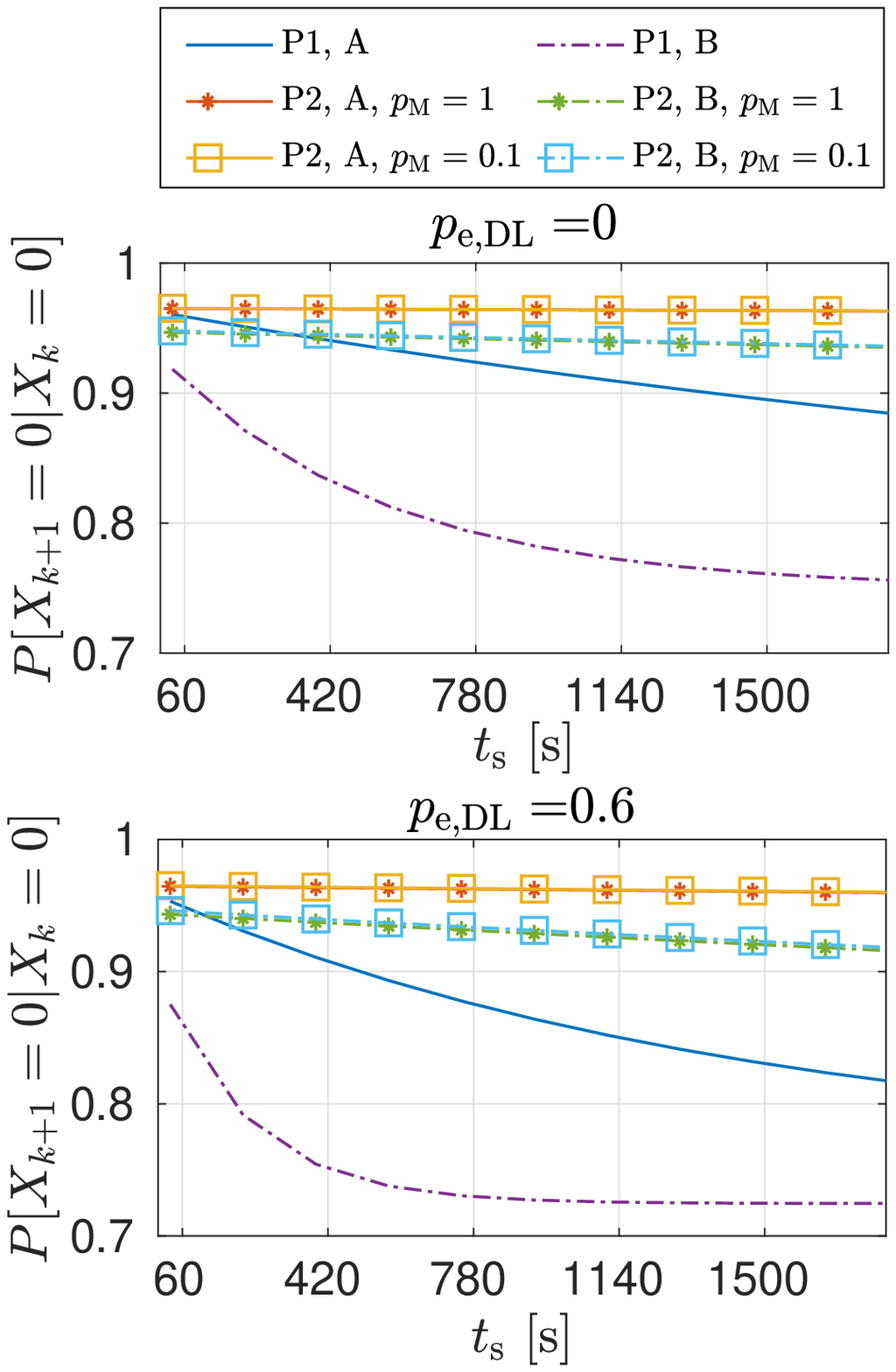}}
  \hfill
  \subfloat[Impact of the block size, protocol P1.]{\includegraphics[width=0.63\columnwidth]{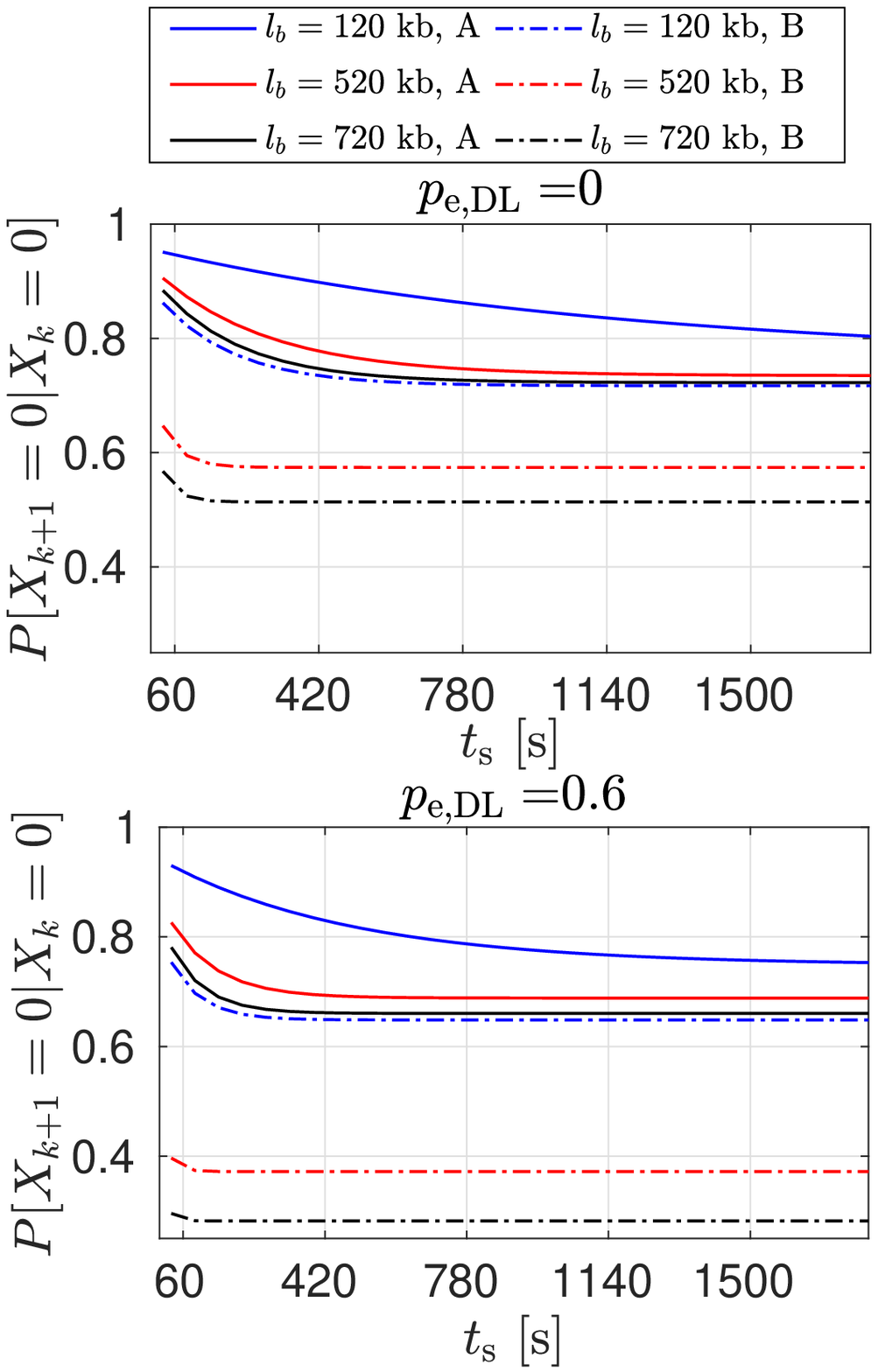}}
  \hfill
  \subfloat[Impact of the block generation frequency, protocol P1.]{\includegraphics[width=0.63\columnwidth]{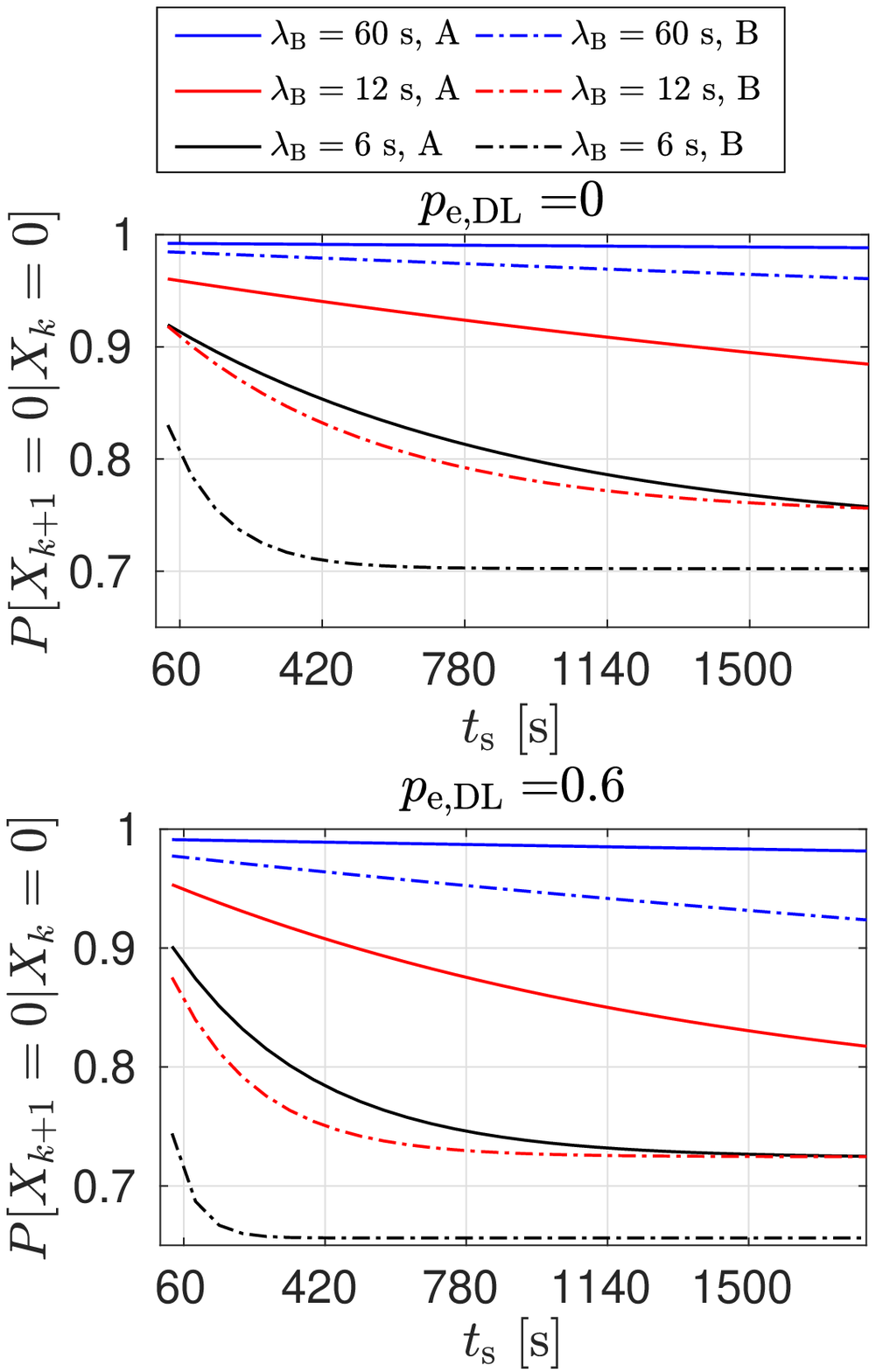}}
    \caption{Probability of staying synchronized when probability of sleeping if fixed to $p_s = 0.2$.}
    \label{fig:p_sync}
\end{figure*}

Fig.~\ref{fig:p_sync} investigates the behavior of probability of staying synchronized, i.e., $P[ X_{k+1} = 0 | X_{k} = 0 ]$, for P1 and P2 and as function of system parameters.
Two cases are considered: error-free communication and a channel that is ideal in uplink but error-prone in downlink; the latter case models the fact that the messages in the downlink for P1 and P2 are much longer than in the uplink.
The probability of sleeping is fixed to $p_{\mathrm{s}} = 0.2$, and we vary the duration of sleeping time $t_{\mathrm{s}}$.
In Fig. \ref{fig:p_sync}(a), we obtain the probability of staying synchronized after the execution of P1 or P2.
The figure shows that, to execute P1, the wireless interface needs to provide an adequate goodput to the blockchain application.
In Figs.~\ref{fig:p_sync}(b) and (c) we vary block size and generation frequency for P1, showing that the probability of staying synchronized is drastically reduced by the sleeping duration, and only slightly by the link unreliability (P2 is not studied here, as it is not affected by variations of the block size).

\begin{figure}[t]
\centering
  \includegraphics[width=0.9\columnwidth]{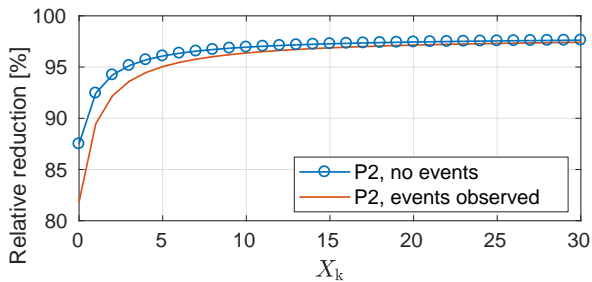}
    \caption{Reduction of downlink data usage in one protocol execution of P2 with respect to the downlink data usage in one protocol execution of P1.}
    \label{fig:data_usage}
\end{figure}

We now investigate the data consumption of the two protocols.
In uplink, protocols are only sending control messages, resulting in a marginal data consumption.
In downlink, the requirements of P1 increase linearly with the accumulated delay, because the full blocks are downloaded, while for P2 they vary with the probability that events of interest are observed, $p_{\mathrm{M}}$.
In Fig.~\ref{fig:data_usage} we show that the downlink data usage is significantly reduced with P2 when compared to P1.

\begin{figure}[t]
\centering
  {\includegraphics[width=1\columnwidth]{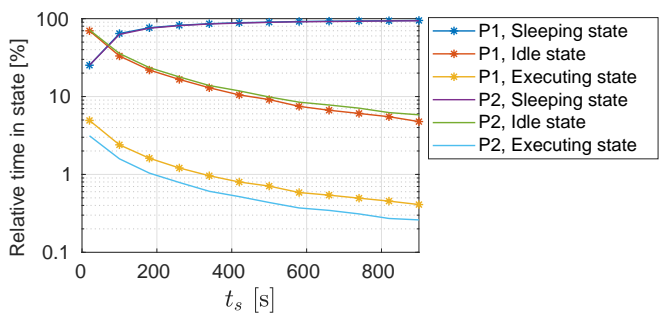}}
    \caption{Fraction of time spent in protocol states.}
    \label{fig:idle}
  \end{figure}

Finally, we show how different sleeping parameters impact the average time spent in the idle, sleeping, and protocol-executing states, see Fig.~\ref{fig:idle}.
We perform a random walk on the Markov Chain corresponding to technology B and sleeping probability $p_\mathrm{s} = 0.2$, while the sleeping duration $t_\mathrm{s}$ is varied.
Note that the fraction of time spent in idle state in P2 is only slightly increased in comparison to P1, although the amount of exchanged data in P2 is significantly smaller than in P1, see In Fig.~\ref{fig:data_usage}.
This is due to the fixed time interval $t_{\mathrm{w}}$ that is present in both protocols, during which a device waits for the response about the current  blockchain state from all $N$ BN nodes to which it is connected (refer to steps $(\mathrm{s}_{1,1})$ and $(\mathrm{s}_{2,1})$ in Sec.~\ref{sec:arch}).

\section{Conclusion}\label{sec:conclusion}

The advent of blockchain-based applications for IoT urges the research on their requirements to the communications system. 
In this work, we investigated several protocols for synchronization between IoT devices and the blockchain network and proposed a model to study the impact of the communication link quality and blockchain parameters on the synchronization process. 
It was showed that the duty cycle of the device should be designed by taking into account both blockchain and communication parameters, as well as that the choice of the wireless technology that guarantees the required reliability of the synchronization process depends on the blockchain parameters.
Also, it was showed that if the protocol execution duration is comparable with the average block-generation period, the probability of staying synchronized rapidly decreases.
In order to reduce the duration of an execution, the use of protocols where the blockchain validation is delegated to external entities is beneficial, although it reduces the devices' security.
Finally, we showed that the blockchain protocols, differently from typical IoT applications that mostly generate uplink traffic, require the allocation of a remarkable amount of downlink resources.

\section*{Acknowledgment}

The work was supported in part by the European Research Council (ERC Consolidator Grant no. 648382 WILLOW) within the Horizon 2020 Program.

%
\IEEEpeerreviewmaketitle



%

\nocite{*}
\bibliographystyle{IEEEtran}
\bibliography{refs_final}

\end{document}